\documentclass[useAMS,usenatbib,onecolumn]{mn2e}

\usepackage{dcolumn}
\usepackage{graphicx}
\usepackage{url}
\usepackage{color}

\newcommand{\cm}{cm$^{-1}$}

\newcommand{\ai}{\textit{ab initio}}

\newcommand{\eqref}[1]{(\ref{#1})}

\newcommand{\p}{^\prime}
\newcommand{\pp}{^{\prime\prime}}

\title{ExoMol line lists II: The ro-vibrational spectrum of SiO}

\date{\today}
\author[Barton et al]{\large Emma J. Barton,
Sergei N. Yurchenko and Jonathan Tennyson \\
Department of Physics and Astronomy, University College London, London WC1E 6BT, UK}
\date{Accepted XXXX. Received XXXX; in original form XXXX}

\pagerange{\pageref{firstpage}--\pageref{lastpage}} \pubyear{2012}

\begin{document}

\maketitle

\begin{abstract}

Accurate rotation-vibration line lists are calculated for silicon
monoxide. Line lists are presented for the main isotopologue,
$^{28}$Si$^{16}$O, and for  four monosubsituted isotopologues
($^{29}$Si$^{16}$O, $^{30}$Si$^{16}$O, $^{28}$Si$^{18}$O and
$^{28}$Si$^{17}$O), in their ground electronic states.
These line lists are suitable for high temperatures (up to 9000 K)
including those relevant to
exoplanetary atmospheres and cool stars.
A combination of empirical and \textit{ab initio} methods is
used: the potential energy curves are determined to high accuracy by
fitting to extensive data from the analysis of both laboratory and sunspot
spectra; a high quality {\it ab initio} dipole moment curve is
calculated at the large basis set, multi-reference configuration
interaction level. A partition function plus full line lists of
rotation-vibration transitions are made available in an electronic
form as supplementary data to this article and at
\url{www.exomol.com}.

\end{abstract}

\label{firstpage}

\section{Introduction}

Silicon monoxide (SiO) is a widely observed astronomical species and
appears to be ubiquitous in our galaxy. Since its
initial detection by \citet{74SnBuxx.SiO}, SiO has proved to be a key astrophysical maser and its  very bright maser emissions
are the subject of continued study \citep{11CoRaDa.SiO,11AsDiRi.SiO,11NaDeIm.SiO,11VlHuFr.SiO,11DeKoKu.SiO}.

SiO was originally detected in the interstellar medium by
\citet{71WiPeJe.SiO}.  Its emissions were subsequently observed in the
envelopes of oxygen-rich giant and super-giant stars
\citep{75KaBuSn.SiO} and the remnants of Supernova 1987A
\citep{88AiSmJa.SiO}. SiO was detected in absorption in cool giant
stars \citep{82RiWixx.SiO} including $\alpha$ Tau
\citep{92CoWiCa.SiO}.  SiO absorptions are prominent in sunspots
which have proved to be a fruitful source of spectroscopic data on highly
excited states \citep{85GlHiJe.SiO,95CaKlDu.SiO} which we exploit
below. Furthermore recent detections of hot, dense exoplanets have
led to the speculation that their atmospheres might contain significant
quantities of SiO from vaporised silicates \citep{12ScLoFe.SiO}.

These astronomical applications, combined with technological uses
of SiO spectra \citep{00WoDaWu.SiO,02MoCoVi.SiO}, have motivated a number
of laboratory studies. \citet{81TiChxx.SiO},\citet{91MoGoVr.SiO},
\citet{95CaKlDu.SiO}, \citet{98ChSaxx.SiO}
and \citet{03SaMcTh.SiO} have all produced molecular constants characterising
SiO vibration-rotation states. Transition probabibilities or Einstein-A
coefficients have also been provided by  \citet{81TiChxx.SiO},
\citet{93LaBaxx.SiO} and \citet{98DrSpEd.SiO}. Transition line lists
have been constructed using these data \citep{81LoMaOl.SiO,85GlHiJe.SiO,93LaBaxx.SiO,98DrSpEd.SiO}, but none of them appears 
to be particularly complete. For example
no single line list combines a comprehensive set of transition frequencies
with an accurate model for the transition intensities. It is this that
we aim to do here.

\citet{81TiChxx.SiO} reported transition probabilities for a large
number of transitions based on a semi-empirical dipole moment which
they claim should give results accurate to about 10~\%.
\citet{93LaBaxx.SiO} computed an accurate electric dipole moment
function for the SiO molecule which not only reproduces accurate
dipole moments, but also reproduces line strengths for vibrational-rotational
transitions within the ground-state manifold. The resulting line lists
are for $^{28}$Si$^{16}$O, $^{29}$Si$^{16}$O and $^{30}$Si$^{16}$O
for $J \leq 250$ but $v$ is limited to 15. Intensities were calculated
using band intensities and H\"onl-London factors. \citet{98DrSpEd.SiO}
calculated an \ai\ $^{28}$Si$^{16}$O line list for $J \leq 100$ and $v
\leq 40$ although the publically available line list from the
Strasbourg Data Centre contains only 503 transitions.
\citet{98DrSpEd.SiO}'s intensities are significantly larger than
those of \citet{81TiChxx.SiO} and they estimate their intensities are
only accurate to about 20~\%.

The ExoMol project \cite{jt528} aims to provide line lists of
spectroscopic transitions for key molecular species which are likely
to be important in the atmospheres of extrasolar planets and cool
stars; it aims, scope and methodology have been summarised by
\citet{jt528}. Line lists for $^2\Sigma^+$ XH molecules, X = Be, Mg,
Ca, have already been published \citep{jt529}.  In the present
paper, we present rotation-vibration transition lists and associated
spectra for the five major isotopologues of SiO. These line lists are
particulary comprehensive and should be valid for temperatures up to
9000 K.

\section{Method}

Rotation-vibration line lists for the ground electronic
state of SiO were obtained by direct solution of the nuclear motion
 Schr\"{o}dinger equation using the program LEVEL8.0 \citep{lr07}. In principle
the calculations were initiated using a potential energy curve (PEC)
calculated {\it ab initio}. In practice, as detailed below, there
are  sufficient experimental data available for SiO that the PEC
was actually characterised by fitting to these data using the program
 DPotFit 1.1 \citep{dpotfit}.

\subsection{Dipole moments}

There appears to be no experimental measurements of any SiO
 transition dipoles and a single measurement of its permanent dipole
 moment as a function of the vibrational state by \citet{70RaMuKl.SiO}. We
 determined a new dipole moment curve (DMC) using high level
 {\it ab initio} calculations. These are compared to previous,
 high-level {\it ab initio} determinations
 \citep{93LaBaxx.SiO,98DrSpEd.SiO,03ChChDa.SiO} given below.

 The \ai\ calculations were performed using MOLPRO \citep{molpro.method};
we tested both the
 coupled cluster (CCSD(T)) and multi-reference configuration
 interaction (MRCI) methods with large basis sets. Our largest, and
 best, calculations used an aug-cc-pCV5Z basis set for the CCSD(T)
 study and aug-cc-pwCV5Z for MRCI; in both cases a correction due to
 the correlation of the core electrons was included but the
 corresponding correction due to relativistic effects was found to be
 very small and was neglected. These calculations give a value for
 the dipole at equilibrium of 3.10 D and 3.07 D, respectively, which
 bracket the experimentally determined value of 3.088 D
 \citep{70RaMuKl.SiO}.  The \ai\ DMC grid points were used directly in
 LEVEL.

\begin{figure}
\begin{center}
\includegraphics{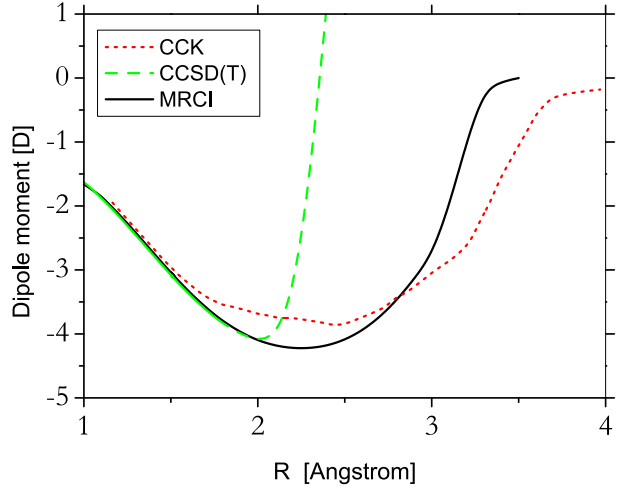}
\caption{{\it Ab initio} dipole moment curves for SiO in its ground electronic state.
MRCI and CCSD(T) are {\it ab initio} calculations performed
as part of this work; CCK is the MRCI dipole of \citet{03ChChDa.SiO}. }
\label{fig:dipole}
\end{center}
\end{figure}

 Figure~\ref{fig:dipole} compares the DMC's arising from our
 calculations with those obtained by \citet{03ChChDa.SiO} (CCK) using
 an MRCI method. Other previous studies based on the use of similar CCSD(T)
  method gave curves similar to our CCSD(T) calculation. The CCSD(T) dipoles appear correct at short bondlengths
 but behave in an unphysical fashion for $R > 2$~\AA. CCSD(T) is known
 to have problems as molecules are dissociated and so this model was not
 pursued. Again at short $R$ our MRCI dipoles are similar to those of
 CCK. It is unclear why CCK's dipoles do not vary smoothly as $R$ is
 increased but such behaviour is not consistent with obtaining reliable
transition intensities. All calculations presented below therefore
use  our MRCI dipole moments.

\begin{table}
\caption{Summary of observations data used to determine the SiO
potential energy curve. Temperatures are only approximate and the
uncertainties are the estimates given in the cited papers.}
\label{tab:obsdata} \footnotesize
\begin{center}
\begin{tabular}{llllll}
\hline
Reference&  Method     &    Transitions&  Temperature & Frequency range    & Uncertainty \\
    &             &               &  (K)         & (cm$^{-1}$)  & (cm$^{-1}$)\\
\hline
\citet{03SaMcTh.SiO} & Laboratory & $\Delta v=0$, $J' \to J" = 1 \to 0$& 1000 -- 9000& 0.13 -- 1.67  & $6.7 \times 10^{-8}$\\
       && $^{28}$Si$^{16}$O $v = 0 - 45$\\
       && $^{29}$Si$^{16}$O $v = 0 - 26$\\
       && $^{28}$Si$^{18}$O $v = 0 - 44$\\
\\
\citet{95CaKlDu.SiO} & Sunspots & $\Delta v=1$, $\Delta J = \pm 1$& 3200& 900 -- 1300  & 0.006\\
       && $^{28}$Si$^{16}$O $v = 0 - 13$ $J \leq 141$\\
       && $^{29}$Si$^{16}$O $v = 0 - 6$ $J \leq 107$\\
       && $^{30}$Si$^{16}$O $v = 0 - 6$ $J \leq 92$\\
\hline
\end{tabular}
\end{center}

\end{table}

\subsection{Fitting the potential}

The potential energy curve is essentially characterised by fitting to
spectroscopic data.  Measurements with very small estimated
uncertainties were chosen for the fit such that any uncertainty in the
fitted surface depends only on the accuracy of the fit and, for higher
energies, the extrapolation beyond the experimental input values.
Measurements for multiple isotopologues were used both to maximise the
input data and to ensure that the resulting curve is applicable to all
isotopologues, including the yet-to-be observed spectrum of
$^{28}$Si$^{17}$O.  The most comprehensive and accurate sets of
available measurements are the infrared ro-vibrational sunspot lines
detected by \citet{95CaKlDu.SiO} and the microwave rotational
laboratory lines recorded by\citet{03SaMcTh.SiO}.

The potential was expressed as an
Extended Morse Oscillator potential function (EMOp):
\begin{equation}
 V(r) = D_{\rm e} \left[ 1 - e^{-\phi(r)\left(r-r_{\rm e}\right)} \right]^2,
\end{equation}
where
\begin{equation}
\phi(r) = \sum_{i=0}^N \phi_i y_p(r,r_{\rm e})^i,
\end{equation}
\begin{equation}
 y_p(r,r_{\rm e}) = \frac{r^p - r_{\rm e}^p}{r^p + r_{\rm e}^p}
\end{equation}
and $p$ was set to 2 and $N$ to 6. Spectroscopically derived values
for the dissociation
energy, $D_{\rm e} = 66620.0$ cm$^{-1}$ \citep{95CaKlDu.SiO} and $r_{\rm e} = 1.5097377$~\AA
\citep{03SaMcTh.SiO} were used.

\begin{table}
\caption{Fitting parameters used in the Extended Morse Oscillator potential,
see eq.~(1). (Uncertainties are given in parentheses in units of the last
digit.)}
\label{tab:potfit} \footnotesize
\begin{center}
\begin{tabular}{ll}
\hline
$N$ & $\phi_i$\\
\hline
0 & 1.86869070(93) \\
1 & -0.142883(30)\\
2 & 0.188883(98)\\
3 & 0.2255(14)\\
4 & 0.1959(37)\\
5 & -0.471(16)\\
6 & 2.714(29)\\
\hline
\end{tabular}
\end{center}
\end{table}

Data for isotopologues were fitted simultaneously; attempts to include
Born-Oppenheimer Breakdown (BOB) terms in the fit did not result in an
improvement and were not pursued.  Of the 1816 lines used in the fit,
1294 were $^{28}$Si$^{16}$O, 301 were $^{29}$Si$^{16}$O, 178 were
$^{30}$Si$^{16}$O and 43 were $^{28}$Si$^{18}$O. These data were used
to determine seven constants with values for $r_{\rm e}$ and $D_{\rm
  e}$ held fixed.  Fits were started from an {\it ab initio} PEC but
were not sensitive to this starting point. The resulting fits
reproduced the input experimental data within 0.01 cm$^{-1}$ and
often, particularly for the pure rotational data, much better than
this.  Parameters resulting from the fit are given in
table~\ref{tab:potfit}.

\subsection{Partition function}

A partition function for $^{28}$Si$^{16}$O was calculated by summing
all the calculated energy levels, see Table~\ref{tab:finallevels}
below, using Excel.  As we use all ro-vibrational energy levels there
are no issues with convergence of this sum.  Given the high accuracy
of our energy levels, this determination should be more accurate than
the previous determinations given by \citep{81Irxxxx.partfunc} and
\citep{84SaTaxx.partfunc}.  Table~\ref{tab:pfcompare} compares with
these previous studies and finds generally good agreement except
the values of \citep{84SaTaxx.partfunc} become too large at higher
temperatures.

\begin{table}
\caption{Comparison of $^{28}$Si$^{16}$O partition functions.}
\label{tab:pfcompare}
\begin{center}
\begin{tabular}{crrr}
\hline
$T$(K) & This work & \citet{84SaTaxx.partfunc} & \citet{81Irxxxx.partfunc}\\
\hline
1000    &  1163.1 &  1169.2  &  1164.2  \\
2000    &  3319.0 &  3316.1  &  3322.0  \\
3000    &  6638.3 &  6618.3  &  6639.8  \\
4000    & 11172.6 & 11000.2  & 11164.1  \\
5000    & 16976.3 & 16629.7  & 16937.5  \\
6000    & 24109.4 & 23833.9  & 24002.6  \\
7000    & 32640.8 & 33080.1  & 32403.5  \\
8000    & 42648.8 & 44989.4  & 42186.6  \\
9000    & 54221.8 & 60371.5  & 53401.1  \\
\hline
\end{tabular}
\end{center}
\end{table}

For ease of use, we fitted our partition function, $Q$, to a series expansion
of the form used by \citet{jt263}:
\begin{equation}
\log_{10} Q(T) = \sum_{n=0}^6 a_n \left[\log T\right]^n \label{eq:pffit}
\end{equation}
with the values given in Table~\ref{tab:pffit}.

\begin{table}
\caption{Fitting parameters used to fit the partition functions,
see eq.~\ref{eq:pffit}. Fits are valid for temperatures between 900 and 9000 K.}
\label{tab:pffit} \footnotesize
\begin{center}
\begin{tabular}{lllllll}
\hline
      &  $^{28}$Si$^{16}$O &  $^{29}$Si$^{16}$O &  $^{30}$Si$^{16}$O   &  $^{28}$Si$^{18}$O    &  \multicolumn{2}{c}{$^{28}$Si$^{17}$O~$^a$}\\
\hline
$a_0$ & -4.289479066  & -2.645361792& -4.290451186     &   -4.296319774     & 9.147577169       & -3.273169177   \\
$a_1$ & 7.19712692    & 4.92053406& 7.19640021         &   7.19145985     &   2.34443076     &   12.27676929    \\
$a_2$ & -1.35646155   & 0.25159971& -1.35602888        &   -1.35513910     &  -12.16129465      &3.08495699   \\
$a_3$ & -1.50671629   & -2.11202057& -1.50490351       &   -1.49684190     &  9.00993487      &  -9.93550466    \\
$a_4$ & 0.909814866   & 1.038226161& 0.909249889       &   0.905530086     &  -2.947037343      &4.767331120  \\
$a_5$ & -0.1829250025 & -0.1975004943& -0.1829028096   &   -0.1823056408     &0.4708553438       &-0.9089004600 \\
$a_6$ & 0.0129066948  & 0.0135985360& 0.0129111795     &   0.0128787601     & -0.0299614057       & 0.0626337008\\
\hline
\end{tabular}

\noindent
  $^a$ Left hand columm, $T \leq 4700$~K; right hand column $T > 4700$~K.
\end{center}
\end{table}

\subsection{Line list calculations}

Line lists were calculated for the five isotopologues
$^{28}$Si$^{16}$O, $^{29}$Si$^{16}$O, $^{30}$Si$^{16}$O,
$^{28}$Si$^{18}$O and $^{28}$Si$^{17}$O. All rotation-vibration states
were considered and transitions satisfying the dipole selection rule
$\Delta J = \pm 1$. A summary of each line list is given in
Table~\ref{tab:finallevels}.
These line lists in principle span frequencies up to 65~000 cm$^{-1}$;
in practice, transitions above 10~000 cm$^{-1}$ are very weak.

The procedure described above was used to produce line lists, i.e. catalogues of transition frequencies $\tilde{\nu}_{ij}$ and
Einstein coefficients $A_{ij}$, for five silicon oxide
isotopologues $^{28}$Si$^{16}$O,
$^{29}$Si$^{16}$O, $^{30}$Si$^{16}$O, $^{28}$Si$^{18}$O and $^{28}$Si$^{17}$O.
The computed line lists for are given in the supplementary materials.

\section{Results}
\begin{table}
\caption{Extract from the state file for $^{28}$Si$^{16}$O. Full tables
are available at http://cdsarc.u-strasbg.fr/cgi-bin/VizieR?-source=J/MNRAS/xxx/yy. The files contain  24~306 entries for $^{28}$Si$^{16}$O,  25~254  for $^{28}$Si$^{17}$O,    26~162  for $^{28}$Si$^{18}$O,  24~617  for $^{29}$Si$^{16}$O and  24~915 for $^{30}$Si$^{16}$O.
 }
\label{tab:levels}
\begin{center}
\begin{tabular}{lrrrrr}
\hline
          $I$  &  $\tilde{E}$      &  $g$  &  $J$ & $v$\\
\hline
           1 &    0.000000 &     1   &    0 &   0\\
           2 &    1.448467 &     3   &    1 &   0\\
           3 &    4.345384 &     5   &    2 &   0\\
           4 &    8.690712 &     7   &    3 &   0\\
           5 &   14.484267 &     9   &    4 &   0\\
           6 &   21.726203 &    11   &    5 &   0\\
\hline
\end{tabular}

\noindent
  $I$:   State counting number;
 $\tilde{E}$: State energy in \cm;
  $g$: State degeneracy;
$J$:   State rotational quantum number;
 $v$:   State vibrational quantum number.

\end{center}
\end{table}

\begin{table}
\caption{Extracts from the transitions file for $^{28}$Si$^{16}$O.
 Full tables
are available from http://cdsarc.u-strasbg.fr/cgi-bin/VizieR?-source=J/MNRAS/xxx/yy. The files contain  1~784~964  entries for $^{28}$Si$^{16}$O, 1~890~039   for $^{28}$Si$^{17}$O,  1~993~414    for $^{28}$Si$^{18}$O, 1~818~923   for $^{29}$Si$^{16}$O and  1~852~656  for $^{30}$Si$^{16}$O.
}
\label{tab:trans}
\begin{center}
\begin{tabular}{ccr}
\hline
       $I$  &  $F$ & $A_{IF}$\\
           2  &          1 & 2.8438E-06\\
           3   &         2  &2.7301E-05\\
           4   &         3 & 9.8720E-05\\
           5   &         4 & 2.4265E-04\\
           6    &         5 & 4.8470E-04\\
           7    &         6 & 8.5037E-04\\
\hline
\end{tabular}

\noindent
 $I$: Upper state counting number;
$F$:      Lower state counting number;
$A_{IF}$:  Einstein A coefficient in s$^{-1}$.

\end{center}
\end{table}

\begin{table}
\caption{Summary of our SiO line lists.}
\label{tab:finallevels}
\begin{center}
\begin{tabular}{lrrrrr}
\hline
& $^{28}$Si$^{16}$O &  $^{29}$Si$^{16}$O &  $^{30}$Si$^{16}$O &$^{28}$Si$^{18}$O & $^{28}$Si$^{17}$O\\
\hline
Maximum $v$ & 95 & 95 & 96 & 98& 97\\
Maximum $J$ & 408 & 410 & 413 & 423 & 416\\
Number of lines & 1784964&1818923&1852656&1993414&1890039\\
\hline
\end{tabular}
\end{center}
\end{table}

The line lists contain almost two million transitions each  and, therefore,
for compactness and ease of use,  divided into separate energy
levels and transitions file. This is done using the standard ExoMol format
\citep{jt548} which is based on a method originally developed for
the BT2 line list \citep{jt378}. Extracts for the start of the $^{28}$Si$^{16}$O files are given in tables~\ref{tab:levels} and \ref{tab:trans}.
 The full
line list for each of these isotopologues can be downloaded from the CDS,
via \url{ftp://cdsarc.u-strasbg.fr/pub/cats/J/MNRAS/xxx/yy}, or
\url{http://cdsarc.u-strasbg.fr/viz-bin/qcat?J/MNRAS//xxx/yy}.
The line lists and
partition function
together with auxiliary data including the potential parameters and
dipole moment functions, as well as the absorption
spectrum given in cross section format
\citep{jt542}, can all be
obtained from there as well as at www.exomol.com.
\begin{table}
\caption{Summary comparison of SiO rotation-vibration line lists. Given are the
isotopologues considered,
the maximum values for vibrational ($v$) and rotational ($J$) states considered,
the maximum change in vibrational state ($\Delta v$), whether intensity information
and a partition function are provided.}
\label{alllists}
\begin{center}
\begin{tabular}{lccccc}
\hline
 Reference                   &\citet{81LoMaOl.SiO}&\citet{85GlHiJe.SiO}& \citet{93LaBaxx.SiO} &\citet{97DrHuSp.SiO}& This work\\
\hline
Isotopes                     & 1              & 3 & 3  & 1 &5 \\
maximum $v$                            &6                    & 10  &  15 & 40 & 98 \\
maximum $J$                            &98                   & 86   & 250 & 1 & 423\\
maximum $\Delta v$& 2                     & 1  & 4 & 3 & 98\\
Intensities? & No                   & No     & Yes & Yes & Yes\\
Partition
Function? & No                   & No      & No & No& Yes\\
\hline
\end{tabular}
\end{center}
\end{table}

\begin{figure}
\begin{center}
\scalebox{0.6}{\includegraphics{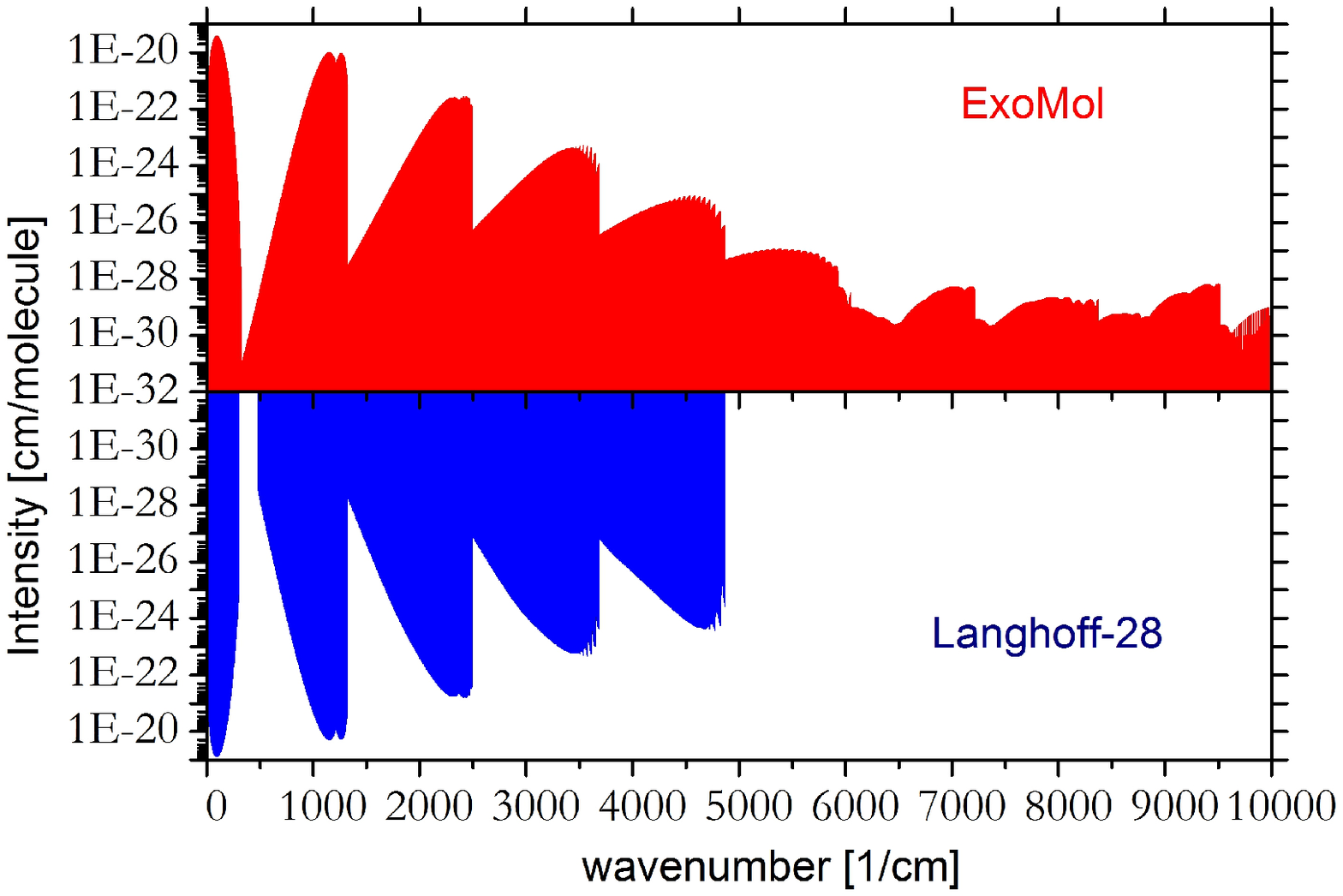}}
\caption{Absorption spectra of $^{28}$Si$^{16}$O  at T=296~K: ExoMol versus
\citep{93LaBaxx.SiO}.}
\label{overview.eps}
\end{center}
\end{figure}

Table~\ref{alllists} compares our SiO line lists with previous attempts
to study this system: it only provides an assessment of the quantity of
data provided not its quality. However it is clear that our new line lists
provide a much more comprehensive coverage of the problem. We believe
they also represent a substantial improvement in accuracy.
Figure~\ref{overview.eps} gives a room temperature comparison of our
$^{28}$Si$^{16}$O line list with the most complete previous one
due to \citet{93LaBaxx.SiO}. It is clear our line list is more
complete.

\begin{table}
\caption{Comparison of theoretically predicted ro-vibrational wavenumbers, in cm$^{-1}$,
with the high-resolution sunspot observations of \citet{95CaKlDu.SiO}. }
\label{tab:SiO:energies} 
\begin{center}
\begin{tabular}{rrrrrrrcrr}
\hline

  $J\p$    & $J\pp$ &   $v\p$    &   $v\pp$  &       Obs.   &      \multicolumn{2}{c}{This work}  && \multicolumn{2}{c}{\citet{93LaBaxx.SiO}}\\ \cline{6-7}\cline{9-10}
           &        &          &           &                 &       Calc.   &  Obs.-Calc.   &&       Calc.   &  Obs.-Calc. \\
\hline
    9      &  10    &     1    &      0    &$     1214.6810 $&$    1214.6829 $&$    -0.0019 $&&$  1214.68 $&$     0.00$\\
    9      &  10    &     2    &      1    &$     1202.8878 $&$    1202.8876 $&$     0.0002 $&&$  1202.89 $&$    -0.00$\\
    9      &  10    &     3    &      2    &$     1191.1305 $&$    1191.1291 $&$     0.0014 $&&$  1191.13 $&$     0.00$\\
    9      &  10    &     4    &      3    &$     1179.4093 $&$    1179.4070 $&$     0.0023 $&&$  1179.41 $&$    -0.00$\\
    9      &  10    &     8    &      7    &$     1132.8751 $&$    1132.8725 $&$     0.0026 $&&$  1132.87 $&$     0.01$\\
    9      &  10    &     10   &      9    &$     1109.8133 $&$    1109.8123 $&$     0.0010 $&&$  1109.81 $&$     0.00$\\
    9      &  10    &     11   &      10   &$     1098.3324 $&$    1098.3329 $&$    -0.0005 $&&$  1098.33 $&$     0.00$\\
    9      &  10    &     13   &      12   &$     1075.4744 $&$    1075.4740 $&$     0.0004 $&&$  1075.48 $&$    -0.01$\\
    11     &  10    &     2    &      1    &$     1232.8767 $&$    1232.8735 $&$     0.0032 $&&$  1232.87 $&$     0.01$\\
    11     &  10    &     3    &      2    &$     1220.9058 $&$    1220.9040 $&$     0.0018 $&&$  1220.91 $&$    -0.00$\\
    11     &  10    &     4    &      3    &$     1208.9734 $&$    1208.9711 $&$     0.0023 $&&$  1208.97 $&$     0.00$\\
    11     &  10    &     5    &      4    &$     1197.0759 $&$    1197.0742 $&$     0.0017 $&&$  1197.08 $&$    -0.00$\\
    11     &  10    &     6    &      5    &$     1185.2130 $&$    1185.2129 $&$     0.0001 $&&$  1185.21 $&$     0.00$\\
    11     &  10    &     7    &      6    &$     1173.3880 $&$    1173.3867 $&$     0.0013 $&&$  1173.39 $&$    -0.00$\\
    11     &  10    &     8    &      7    &$     1161.5964 $&$    1161.5954 $&$     0.0010 $&&$  1161.60 $&$    -0.00$\\
    11     &  10    &     10   &      9    &$     1138.1158 $&$    1138.1155 $&$     0.0003 $&&$  1138.12 $&$    -0.00$\\
    10     &  11    &     1    &      0    &$     1213.1327 $&$    1213.1350 $&$    -0.0023 $&&$  1213.13 $&$     0.00$\\
    10     &  11    &     2    &      1    &$     1201.3510 $&$    1201.3499 $&$     0.0011 $&&$  1201.35 $&$     0.00$\\
    10     &  11    &     3    &      2    &$     1189.6031 $&$    1189.6015 $&$     0.0016 $&&$  1189.60 $&$     0.00$\\
    10     &  11    &     5    &      4    &$     1166.2155 $&$    1166.2135 $&$     0.0020 $&&$  1166.21 $&$     0.01$\\
    10     &  11    &     6    &      5    &$     1154.5742 $&$    1154.5727 $&$     0.0015 $&&$  1154.57 $&$     0.00$\\
    10     &  11    &     7    &      6    &$     1142.9704 $&$    1142.9668 $&$     0.0036 $&&$  1142.97 $&$     0.00$\\
    10     &  11    &     9    &      8    &$     1119.8600 $&$    1119.8586 $&$     0.0014 $&&$  1119.86 $&$     0.00$\\
    10     &  11    &     12   &      11   &$     1085.4509 $&$    1085.4502 $&$     0.0007 $&&$  1085.45 $&$     0.00$\\
    10     &  11    &     13   &      12   &$     1074.0547 $&$    1074.0474 $&$     0.0073 $&&$  1074.05 $&$     0.00$\\
\\
   125     &  126   &     2    &     1    &$      965.1065 $&$     965.1051 $&$     0.0014 $&&$   965.09 $&$     0.02$\\
   125     &  126   &     3    &     2    &$      954.5828 $&$     954.5856 $&$    -0.0028 $&&$   954.57 $&$     0.01$\\
   125     &  126   &     4    &     3    &$      944.0942 $&$     944.0980 $&$    -0.0038 $&&$   944.09 $&$     0.00$\\
   126     &  127   &     1    &     0    &$      973.1300 $&$     973.1257 $&$     0.0043 $&&$   973.11 $&$     0.02$\\
   126     &  127   &     2    &     1    &$      962.5855 $&$     962.5854 $&$     0.0001 $&&$   962.57 $&$     0.02$\\
   126     &  127   &     4    &     3    &$      941.5952 $&$     941.6004 $&$    -0.0052 $&&$   941.59 $&$     0.01$\\
   127     &  128   &     1    &     0    &$      970.5914 $&$     970.5876 $&$     0.0038 $&&$   970.57 $&$     0.02$\\
   127     &  128   &     2    &     1    &$      960.0589 $&$     960.0585 $&$     0.0004 $&&$   960.04 $&$     0.02$\\
   127     &  128   &     3    &     2    &$      949.5598 $&$     949.5612 $&$    -0.0014 $&&$   949.55 $&$     0.01$\\
   127     &  128   &     4    &     3    &$      939.0904 $&$     939.0956 $&$    -0.0052 $&&$   939.08 $&$     0.01$\\
   128     &  129   &     1    &     0    &$      968.0462 $&$     968.0425 $&$     0.0037 $&&$   968.03 $&$     0.02$\\
   128     &  129   &     2    &     1    &$      957.5253 $&$     957.5244 $&$     0.0009 $&&$   957.51 $&$     0.02$\\
   128     &  129   &     3    &     2    &$      947.0347 $&$     947.0382 $&$    -0.0035 $&&$   947.02 $&$     0.01$\\
   129     &  130   &     2    &     1    &$      954.9832 $&$     954.9833 $&$    -0.0001 $&&$   954.97 $&$     0.01$\\
   130     &  131   &     1    &     0    &$      962.9348 $&$     962.9306 $&$     0.0042 $&&$   962.91 $&$     0.02$\\
   130     &  131   &     2    &     1    &$      952.4351 $&$     952.4349 $&$     0.0002 $&&$   952.42 $&$     0.02$\\
   130     &  131   &     3    &     2    &$      941.9681 $&$     941.9710 $&$    -0.0029 $&&$   941.95 $&$     0.02$\\
   131     &  132   &     1    &     0    &$      960.3668 $&$     960.3641 $&$     0.0027 $&&$   960.35 $&$     0.02$\\
   131     &  132   &     2    &     1    &$      949.8787 $&$     949.8795 $&$    -0.0008 $&&$   949.86 $&$     0.02$\\
   131     &  132   &     3    &     2    &$      939.4210 $&$     939.4266 $&$    -0.0056 $&&$   939.41 $&$     0.01$\\
   132     &  133   &     1    &     0    &$      957.7940 $&$     957.7905 $&$     0.0035 $&&$   957.77 $&$     0.02$\\
   132     &  133   &     2    &     1    &$      947.3225 $&$     947.3171 $&$     0.0054 $&&$   947.30 $&$     0.02$\\
   132     &  133   &     3    &     2    &$      936.8721 $&$     936.8753 $&$    -0.0032 $&&$   936.85 $&$     0.02$\\
   134     &  135   &     2    &     1    &$      942.1713 $&$     942.1709 $&$     0.0004 $&&$   942.15 $&$     0.02$\\
   135     &  136   &     2    &     1    &$      939.5876 $&$     939.5874 $&$     0.0002 $&&$   939.56 $&$     0.03$\\
   136     &  137   &     1    &     0    &$      947.4301 $&$     947.4255 $&$     0.0046 $&&$   947.40 $&$     0.03$\\
   137     &  138   &     1    &     0    &$      944.8216 $&$     944.8167 $&$     0.0049 $&&$   944.79 $&$     0.03$\\
   137     &  138   &     2    &     1    &$      934.3987 $&$     934.3992 $&$    -0.0005 $&&$   934.37 $&$     0.03$\\
   139     &  140   &     2    &     1    &$      929.1778 $&$     929.1832 $&$    -0.0054 $&&$   929.15 $&$     0.03$\\
   140     &  141   &     1    &     0    &$      936.9546 $&$     936.9489 $&$     0.0057 $&&$   936.92 $&$     0.03$\\
   140     &  141   &     2    &     1    &$      926.5650 $&$     926.5649 $&$     0.0001 $&&$   926.53 $&$     0.04$\\
\hline
\end{tabular}
\end{center}
\end{table}

Illustrative high resolution comparisons with $^{28}$Si$^{16}$O
observed sunspot transition frequencies are given in
Table~\ref{tab:SiO:energies}; as can be seen the comparison
is excellent.  Also given are the frequencies due to
\citet{93LaBaxx.SiO}, who generated a PEC from the empirical parameters of \citet{85GlHiJe.SiO},
again the comparison is very good. Overall the error in Langhoff and Bauschlicher's frequencies, where
available, is only about twice ours.

An important aim in fitting the PEC is to also predict spectroscopic
data for higher vibrational states than used in the fit. To test this
comparisons are made with the experimentally-derived vibrational frequencies
of \citet{76ShLiVe.SiO}, which have an estimated uncertainty 1.0
cm$^{-1}$, and \citet{73BrCoDu.SiO}, as re-assigned by
\citet{75BaStxx.SiO}.  Our predictions agree well with the
experimentally-derived values; however those of \citet{81TiChxx.SiO} (TC)
disagree siginificantly for the higher vibrational states.  Indeed by
$v=35$ our vibrational term value is 3.3 cm$^{-1}$ lower than TC's.
TC were limited in the vibrational states they could use in their fit,
extending only to $v = 5$ from the spectra of \citet{81LoMaOl.SiO},
compared to the line frequencies for vibrational transitions used
which extend to $v = 13$.

\begin{table}
\caption{ Comparison of theoretically predicted vibrational spacings, in cm$^{-1}$,
with the low-resolution
experiments of \citet{73BrCoDu.SiO} for $13 \leq v \leq 17$, and  \citet{73BrCoDu.SiO}, as re-assigned by \citet{75BaStxx.SiO}, for $23 \leq v \leq 28$.}
\label{tab:BeH:energies} \footnotesize
\begin{center}
\begin{tabular}{lrrrr}
\hline
$v' - v``$ &  Obs.     &   Calc.&  \multicolumn{2}{c}{Observed $-$ Calculated} \\ \cline{4-5}
           &      & This work & This work & \citet{81TiChxx.SiO}\\
         \hline
14 - 13     &1078.5      &1077.72   &0.8      & 0,8      \\
15 - 14     &1067.0      &1066.27   &0.7     &0.7      \\
16 - 15     &1055.0      &1054.86   &0.1     &0.1     \\
17 - 16     &1044.0      &1043.48   &0.5      &0.4     \\
18 - 17     &1033.0      &1032.13   &0.9     &0.8      \\
24 - 23     &964.69      &964.70   &-0.01     &-0.36     \\
25 - 24     &953.55      &953.58    &-0.03     &-0.04     \\
26 - 25     &942.28      &942.48    &-0.20     &-0.03     \\
27 - 26     &931.37      &931.41    &-0.04     &-0.52     \\
28 - 27     &920.22      &920.38    &-0.16     &-0.74 \\
\hline
\end{tabular}
\end{center}
\end{table}

\begin{figure}
\begin{center}
\scalebox{0.6}{\includegraphics{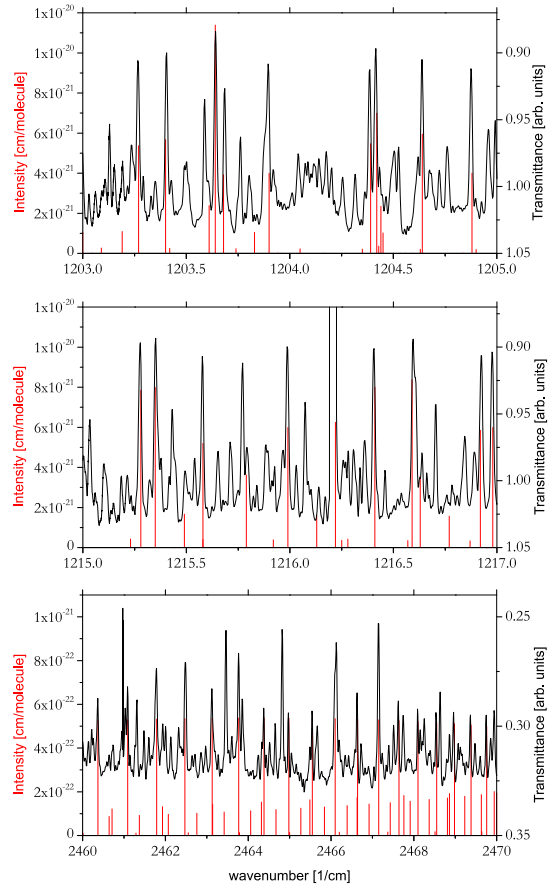}}
\caption{
Sunspot absorption spectra \protect\citep{96WaLiHi.db}, continuous line and our predicted
$^{28}$Si$^{16}$O absorptions at 3200~K, vertical lines.
The absolute intensity scale refers to the calculated spectra.
}
\label{sun.eps}
\end{center}
\end{figure}

Since
there are no  measured intensities to compare with, we
compare the results of our calculations with astronomical spectra.
Sunspots display many SiO features over a wide range of wavelengths.
High resolution sunspot spectra have been compiled by
\citet{96WaLiHi.db}. Figure~\ref{sun.eps} compares small regions of
these spectra with our line lists.  Note that only the theoretical
spectra, generated at $T$=3200~K, are given in the absolute units of
cm$/$molecule).  These spectra, which are chosen to cover regions
where $\Delta v$ is both 1 and 2, illustrate the excellent agreement
between our line positions and the observed features.  Our intensities
are also well correlated with the observed ones. We note that the
sunspot spectra show many other features in this figure.  These
features are either due to water \citep{jt200} or so to some so-far
unidentified species.

\begin{figure}
\begin{center}
\scalebox{0.4}{{\includegraphics{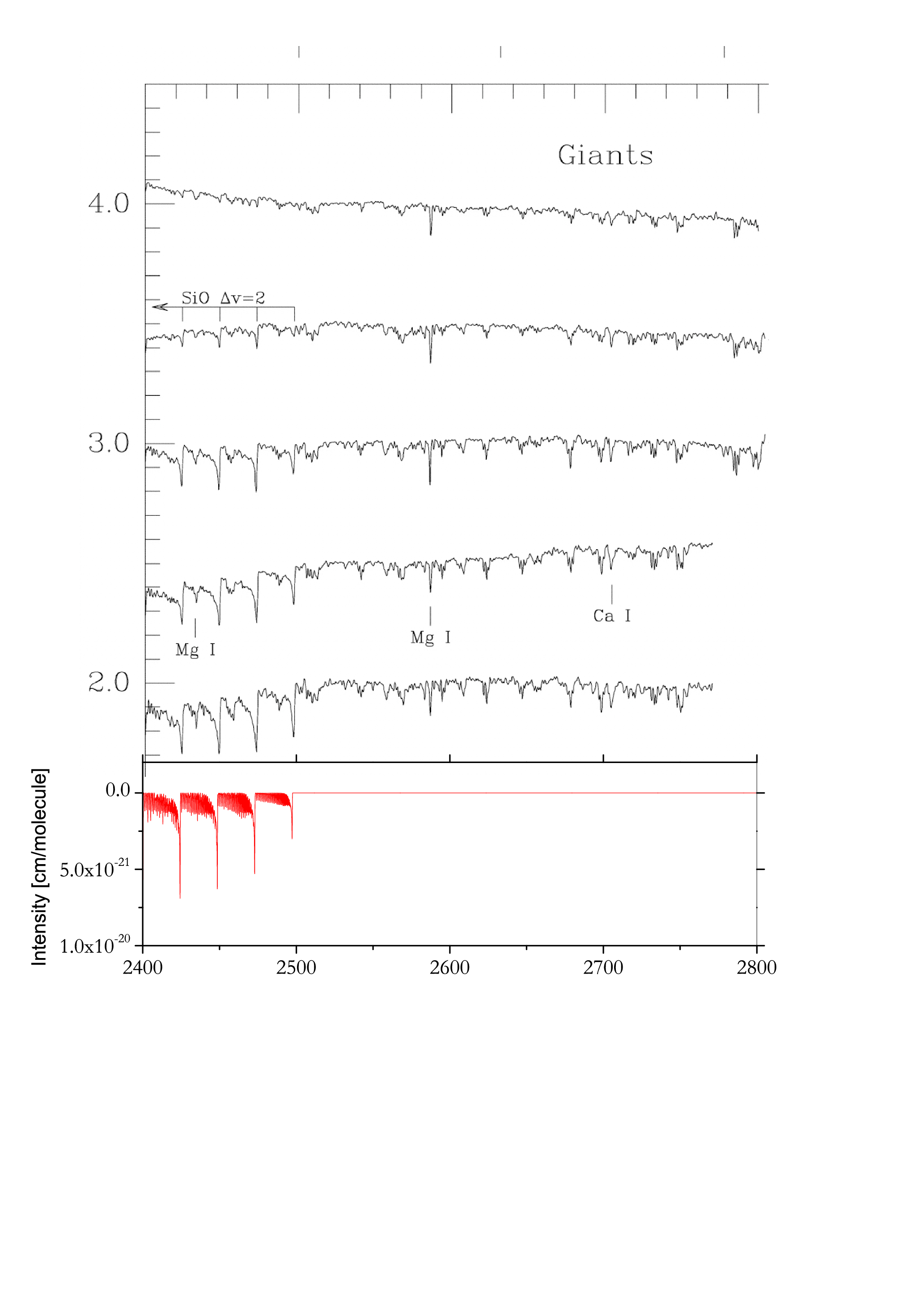}}}
\caption{Red giant spectra due to \citet{02WaHixx.SiO} compared to a
synthesized $^{28}$Si$^{16}$O  absorption spectrum using $T=$3200~K (bottom panel).}
\label{giant.eps}
\end{center}
\end{figure}

Figure~\ref{giant.eps} compares the spectra of four
red giant \citet{02WaHixx.SiO}
with our simulated spectrum of  $^{28}$Si$^{16}$O at a temperature
of 3200~K. These spectra are for the $\Delta v =2$ overtone spectrum. To fully
reproduce the observed stellar spectrum would require running a stellar
model which is beyond the scope of this paper. However it clear that both
the line positions and general band structure predicted by our
calculations are in very good agreement with the observations.

\section{Conclusions}

We present comprehensive line lists for the five most important
isotopologues of SiO. These are based on the direct solution of the
nuclear motion Schr\"odinger equation using a potential energy curve
obtained by fitting to an extensive data set of measured transitions.
These data are reproduced to near experimental accuracy resulting in
high accuracy line positions. A new {\it ab initio} dipole moment is
computed, which appears to behave more physically at large
internuclear separations. This dipole is used to compute Einstein A
coefficients for all possible dipole-allowed transitions within each
SiO isotopologue. The result is a comprehensive line list for each
species, including the first data for $^{28}$Si$^{17}$O. The line
lists can be downloaded from the CDS, via
ftp://cdsarc.u-strasbg.fr/pub/cats/J/MNRAS/, or from
http://cdsarc.u-strasbg.fr/viz-bin/qcat?J/MNRAS/, or from
www.exomol.com.

\section*{Acknowledgement}

This work is supported by ERC Advanced Investigator Project 267219.

\bibliographystyle{mn2e}

\label{lastpage}

\end{document}